\documentclass[%
pre,
superscriptaddress,
preprint,
preprintnumbers,
 amsmath,amssymb,
 aps,
]{revtex4-2}

\usepackage{graphicx}
\usepackage{dcolumn}
\usepackage{bm}
\usepackage{todonotes}
\usepackage{caption}
\usepackage{hyperref}

\hypersetup{
    colorlinks,
    linkcolor={blue},
    citecolor={blue},
    urlcolor={blue}
}

\begin{document}

\title{Modeling the influence of interactions on different variables in a turbulent thermoacoustic system}

\author{Aneesh Srivatsa}
\altaffiliation[presently at ]{Department of Aerospace Engineering, University of Michigan, Ann Arbor, Michigan 48109, USA}
\affiliation{Centre of Excellence for studying Critical Transitions in Complex Systems, Indian Institute of Technology Madras, Chennai 600036, India}
\affiliation{Department of Aerospace Engineering, Indian Institute of Technology Madras, Chennai 600036, India}
\author{Shruti Tandon}%
 \email{shrutitandon97@gmail.com}
 \affiliation{Centre of Excellence for studying Critical Transitions in Complex Systems, Indian Institute of Technology Madras, Chennai 600036, India}
 \affiliation{Department of Aerospace Engineering, Indian Institute of Technology Madras, Chennai 600036, India}
\author{Andrea Elizabeth Biju}
\altaffiliation[presently at ]{Department of Materials science and Mechanical engineering, Harvard University, Cambridge, Massachusetts 02138, USA}
\affiliation{Centre of Excellence for studying Critical Transitions in Complex Systems, Indian Institute of Technology Madras, Chennai 600036, India}
\affiliation{Department of Aerospace Engineering, Indian Institute of Technology Madras, Chennai 600036, India}
\author{Norbert Marwan}
\affiliation{Potsdam Institute for Climate Impact Research (PIK), Member of the Leibniz Association, 14412 Potsdam}
\affiliation{Institute of Geosciences, University of Potsdam, Potsdam 14476, Germany}
\author{J\"urgen Kurths}
\affiliation{Potsdam Institute for Climate Impact Research (PIK), Member of the Leibniz Association, 14412 Potsdam}
\affiliation{Institute of Physics, Humboldt Universit\"at zu Berlin, Berlin 12489, Germany}
\author{R. I. Sujith}
\affiliation{Centre of Excellence for studying Critical Transitions in Complex Systems, Indian Institute of Technology Madras, Chennai 600036, India}
\affiliation{%
Department of Aerospace Engineering, Indian Institute of Technology Madras, Chennai 600036, India
}

\maketitle

{Turbulent reacting flows confined to ducts are plagued by thermoacoustic instability, a state in which a positive feedback between flow, flame and acoustic perturbations leads to the emergence of catastrophically high-amplitude oscillatory dynamics in the sound and global heat release rate fluctuations. Modeling the interdependence between local interactions and the global emergence of order in such spatially extended complex systems is exacting. Here, we present a novel reduced-order model to capture the influence of the local interactions on the variables exhibiting global emergence of order in a turbulent reacting flow system. We represent each variable that exhibits global oscillatory instability as an oscillator with a cubic nonlinearity. The oscillator is driven by a forcing term that represents the holistic influence of the inter-subsystem interactions on the global behavior. The forcing term essentially couples the local interactions and the globally emergent dynamics in the model. Further, the influence of the inter-subsystem interactions on the behavior of each subsystem is different. Therefore, we use different forcing terms for each variable inspired by the physical interactions in the system. The nonlinear oscillators representing the acoustic and the heat release rate oscillations are hence forced using Wiener and Markov-modulated Poisson processes, respectively. Using this approach, we are able to reproduce (i) the multifractal characteristics of acoustic pressure fluctuations during chaotic dynamics, (ii) the loss of multifractality through the experimentally observed scaling law behavior during the transition from chaos to order and (iii) the emergence of periodicity and bifurcation in heat release rate dynamics.
}
\newpage
\section{Introduction}

Complex systems theory describes emergent dynamics that arise from the rich interactions between constituent elements \cite{ladyman2013complex}, that the mere superposition of the behavior of the elements cannot explain. Instead of analyzing individual interacting elements independently, complex systems theory advocates studying the system as a whole by accounting for the holistic effect of their interactions \cite{siegenfeld2020introduction}. Examples of complex systems are found across disciplines such as in biology \cite{jensen1998self}, social science \cite{sawyer2005social}, climatology \cite{tsonis2006networks}, earth systems\cite{fan2021statistical}, and engineering\cite{sujith2021thermoacoustic}. The behavior of these systems is indeed difficult to predict due to the inherent nonlinearity and the large web of interactions, hence, making it challenging to create accurate mathematical models. We adopt a novel perspective to devise a reduced-order model of the emergent dynamics of complex systems.

Here, we present a model for the emergent dynamics in turbulent thermoacoustic systems, such as those in rocket and gas-turbine engines, which are essentially complex systems \cite{sujith2021thermoacoustic}. There, turbulent reacting flow is confined inside a closed duct and interacts with the acoustic field established in the duct. Such systems exhibit a transition from low-amplitude chaotic to high-amplitude self-sustained periodic dynamics in the acoustic pressure signal\cite{ananthkrishnan2005reduced,lieuwen2002experimental} through intermittency dynamics \cite{nair2014intermittency,kabiraj2012nonlinear,delage2017off,guan2020intermittency}. Also, the fluctuations in the global heat release rate signal become periodic and exhibit generalized synchronization with the fluctuations in the acoustic pressure signal during the state of ordered dynamics\cite{pawar2017thermoacoustic}. Such emergence of order is undesirable to engineers, as this self-sustained periodic dynamics often have ruinously large-amplitude that can lead to structural failures, overwhelm the thermal protection systems and destroy mechanical components of the engine leading to mission failures in rockets and huge economic losses in gas-turbine engines \cite{lieuwen_2012}. 

\begin{figure}[h!]
    \centering
    \includegraphics[width=0.7\linewidth]{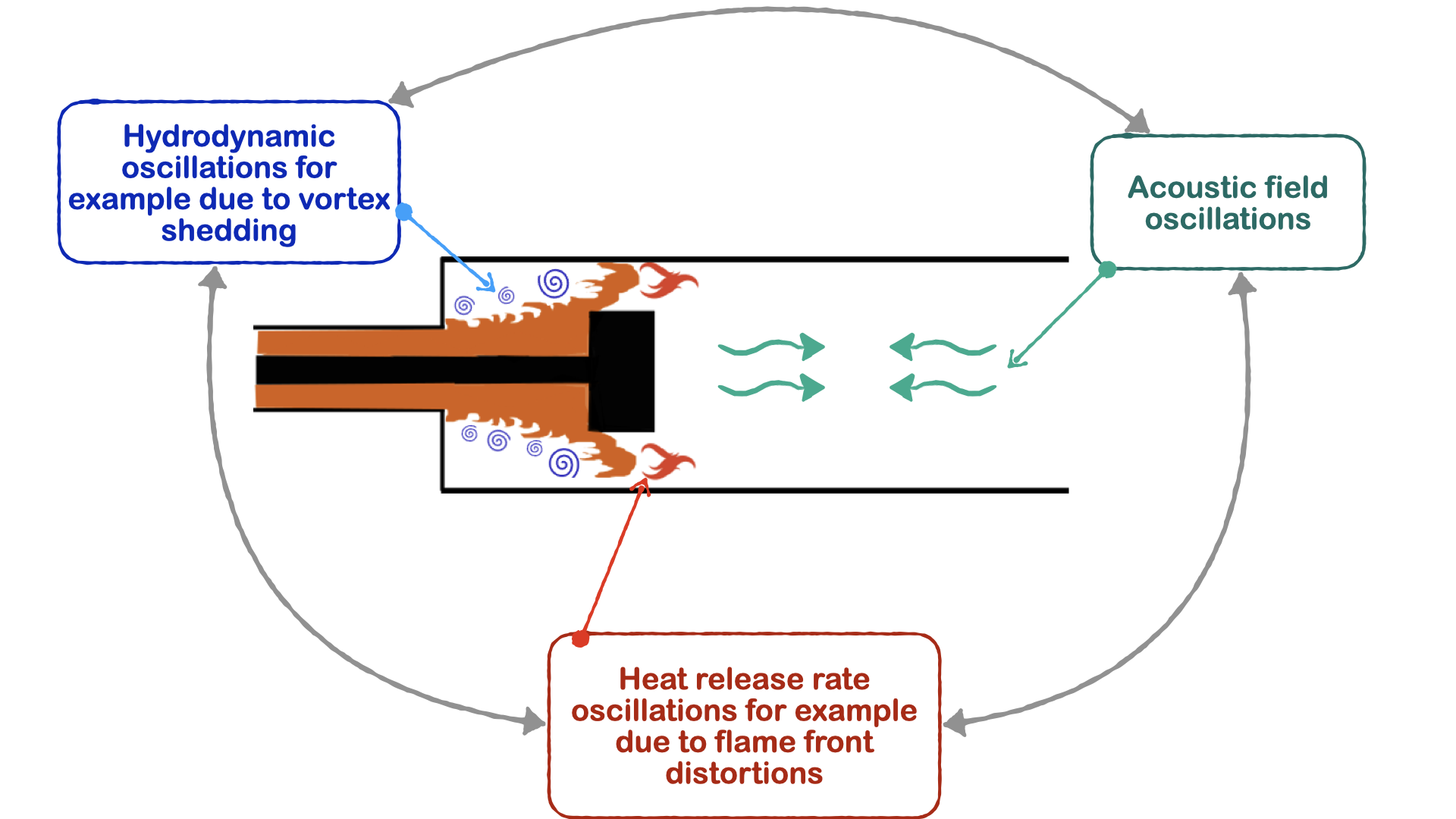}
    \caption{\textbf{Flame-flow-acoustic interactions in turbulent thermoacoustic systems result in emergent order in the acoustic pressure.} The schematic shows the interactions between the various subsystems in a turbulent thermoacoustic system. The configuration of the system includes a backward-facing step that creates a flow recirculation zone due to the sudden expansion in the cross section area. The vortices that are shed from the backward-facing step convect downstream and impinge on the bluff body surface. The mixture of hot fuel and air contained in these vortices undergoes combustion, causing a sudden spike (‘kick’\cite{matveev2003model}) in the heat release rate fluctuations. Acoustic waves are generated in the duct and interact with the local flame and flow fluctuations thus affecting the combustion and vortex shedding processes. Thus, all the subsystems interact in a closed feedback loop.}
    \label{fig_scehmatic_feedback}
\end{figure}

We study the dynamics of turbulent thermoacoustic systems in a turbulent combustor where a bluff-body is used to stabilize and hold the turbulent flame against the underlying turbulent flow. In our system, turbulent flow moves past a sudden area expansion (backward-facing step) and a bluff-body creating recirculation zones. Hot products from combustion are retained in these zones allowing ignition of the incoming reactants. Recirculation zones are associated with the formation of coherent circulatory flow structures called vortices that shed and convect downstream. The size and shedding time of these vortices depends on the acoustic pressure perturbations which, in turn, depend on the energy in the acoustic field \cite{matveev2003model, seshadri2016reduced, nair2015reduced}. Moreover, the energy in the acoustic field is derived from the mean flow through the interactions between the acoustic and heat release rate fluctuations. The fluctuations in the heat release rate occur due to flame distortion by the vortices. Clearly, the vorticity, heat release rate and acoustic dynamics are all interdependent and involved in a closed feedback loop \cite{sujith2020complex} as depicted in Fig. \ref{fig_scehmatic_feedback}. 

Furthermore, these interactions occur in a spatially extended domain and become organized in local pockets of the flow during the emergence of order in the system \cite{krishnan2019emergence, krishnan2021suppression, tandon2023multilayer}. Understanding the temporal features and spatial patterns of flame-acoustic interactions is crucial to predict and control thermoacoustic instability. The flame-acoustic interdependence in thermoacoustic systems is an intrinsic process in the system and can be viewed as a multi-scale interaction between acoustic and heat release rate fluctuations \cite{tandon2020bursting}. Further, a unique spatial pattern of amplitude-frequency coupling occurs between acoustic and heat release rate fluctuations during the occurrence of intermittency prior to the ordered state \cite{tandon2025cross}. Despite the presence of global synchrony between acoustic pressure fluctuations and heat release rate fluctuations \cite{pawar2017thermoacoustic}, local interactions of these fluctuations in the reaction field lead to frequency desynchrony in small spatial pockets during the ordered state \cite{pawar2024spatiotemporal}. How the local interactions between acoustic and heat release rate oscillations influence the features of each of these signals is yet to be understood. 

We note that the dynamical transition from chaos to order in turbulent thermoacoustic systems presents a unique challenge to modeling attempts. Firstly, the system is high-dimensional and involves interactions between subsystems in a spatially extended field and across multiple timescales associated which each subsystem. Moreover, the route of transition obeys a scaling law relation between the amplitude and Hurst exponent (a measure of multifractality explained in Appendix \ref{app_Estimation_hurst}) of acoustic pressure oscillations, and this scaling law is universally obeyed in other turbulent fluid systems such as aero-acoustic and aeroelastic systems\cite{pavithran2020universality}. This scaling behavior is observed in the acoustic pressure signal along with the loss of multifractality due to the emergence of periodicity \cite{nair2014multifractality}. Developing a conceptual reduced-order model for the global oscillating variables, while accounting for the main qualitative and quantitative features of the transition, is essential for understanding and predicting the emergence of order in turbulent thermoacoustic systems. 

Recently, \textcite{weng2023synchronization} replicated the universal scaling law \cite{pavithran2020universality} using a model, where the heat release rate dynamics is represented by a pair of co-evolving second-order differential equations coupled to a single nonlinear oscillator representing the acoustic pressure oscillations. In another study, \textcite{garcia2024universality} used the complex Ginzburg-Landau equation (CGLE) with a global coupling term to model the transition from chaos to order as a transition from defect to phase turbulence in a set of diffusively coupled Stuart-Landau oscillators and recovered the universal scaling law observed in experiments \cite{pavithran2020universality}. Clearly, the quantitative features of the chaos-to-order transition can be replicated if the nonlinearity and inter-subsystem coupling are incorporated into the model. Additional investigation is necessary to determine if the two co-evolving signals, namely the acoustic and the heat release rate signals are affected  similarly by inter-subsystem interactions in the system.

Both the acoustic pressure ($p'$) and heat release rate signals ($\dot{q}'$) are two distinct variables measured in the same system during a certain dynamical state. In fact, we can consider both $p'$ and $\dot{q}'$ as projections of the original system dynamics into a low-dimensional space that occurs in a high-dimensional space. Then, the underlying nonlinearity of both these signals should be the same. Here, we present a model where two identical oscillators are used to represent the $p'$ and $\dot{q}'$ dynamics respectively. Both these oscillators are modeled as Stuart-Landau oscillators that are governed by first order differential equations with a cubic nonlinearity. Note that the bifurcation exhibited by a turbulent thermoacoustic system is similar to a Hopf bifurcation as  the system transitions to limit cycle oscillations \cite{lieuwen2002experimental, ananthkrishnan2005reduced}. However, the transition to limit cycle oscillations is preceded by a state of intermittency in turbulent thermoacoustic systems \cite{nair2014intermittency}. The Stuart-Landau oscillator is the simplest nonlinear oscillator that undergoes a Hopf bifurcation, while preserving phase invariance \cite{garcia2012complex}. Stuart-Landau oscillators have been used extensively to model a variety of oscillatory dynamics, such as the oscillating wake of a cylinder\cite{le2001hysteresis, olinger1993low} and a bluff body under periodic forcing\cite{herrmann2020modeling}, and also rhythmic neural activity \cite{doelling2021neural, powanwe2021amplitude}.

Even though the underlying nonlinearity governing both $p'$ and $\dot{q}'$ is the same, the effective influence of the inter-subsystem interactions on the dynamics of each variable can be very different. As a result, the `topological features' \cite{estrada2023complex} (such as amplitude or frequency modulations) of the observed variables are distinct. For example, the heat release rate signal exhibits periodic spikes, called `kicks', during the state of thermoacoustic instability due to periodic shedding of fuel-laden vortices and sudden combustion thereafter \cite{matveev2003model, seshadri2016reduced}. In fact, \textcite{pawar2017thermoacoustic} used a sine of sine signal ($sin(\omega t + sin(\omega t + \phi_2) +\phi_1)$) to qualitatively mimic the spiky behavior of the $\dot{q}'$ time series during the state of thermoacoustic instability. On the other hand, the acoustic pressure signal exhibits a clean sinusoidal variation during this state.

We note that the observables of a complex system represent the dynamics of a complex system. Here, our aim is to develop a parsimonious representation of the nonlinearity and the influence of interactions on the global oscillatory variables in our system. To do so, we must be able to replicate the most prominent characteristic features of the global observables, namely, $p'$ and $\dot{q}'$. As noted by \textcite{estrada2023complex}, an attempt to model the topological features of the observables, can help infer some of the prominent causal physical processes in a complex system. Here, we use two different forcing terms on the two Stuart-Landau oscillators representing the $p'$ and $\dot{q}'$ dynamics. Note that, even though a forcing term is introduced, the $p'$ and $\dot{q}'$ are co-evolving, interdependent and self-sustained oscillations. Thus, the distinct forcing terms should be interpreted as a representation of the unique holistic influence of flame-fluid-acoustic interactions on each of these variables. 

For the Stuart-Landau oscillator representing the acoustic pressure signal, a forcing by noise generated by the Wiener process \cite{levy2020wiener} is able to capture the qualitative features of the signal during various dynamical states as well as the universal scaling observed for $p'$ in experiments. Further, we use a doubly stochastic process, specifically a Markov modulated Poisson process (MMPP)\cite{fischer1993markov}, to force the Stuart Landau oscillator representing $\dot{q}'$. Markov modulated Poisson processes have found many interesting applications such as in modeling criminal intrusion in internet or financial networks \cite{scott2001detecting}, population arrival dynamics \cite{ihler2006adaptive}, multimedia and internet traffic \cite{shah2000mmpp, muscariello2005markov}, and occurrence of deep earthquakes\cite{lu2012markov}. The motivation behind each of these forcing terms is elaborated in Sec. \ref{Methods}. In Sec. \ref{experimental setup} we discuss the setup of a laboratory-scale turbulent thermoacoustic system and the experiments performed. The results from the model presented in Sec. \ref{Methods} are discussed and critically compared with the experiments in Sec. \ref{Results}.

\section{Experimental setup} \label{experimental setup}

\begin{figure}[h!]
    \centering
    \includegraphics[width=\linewidth]{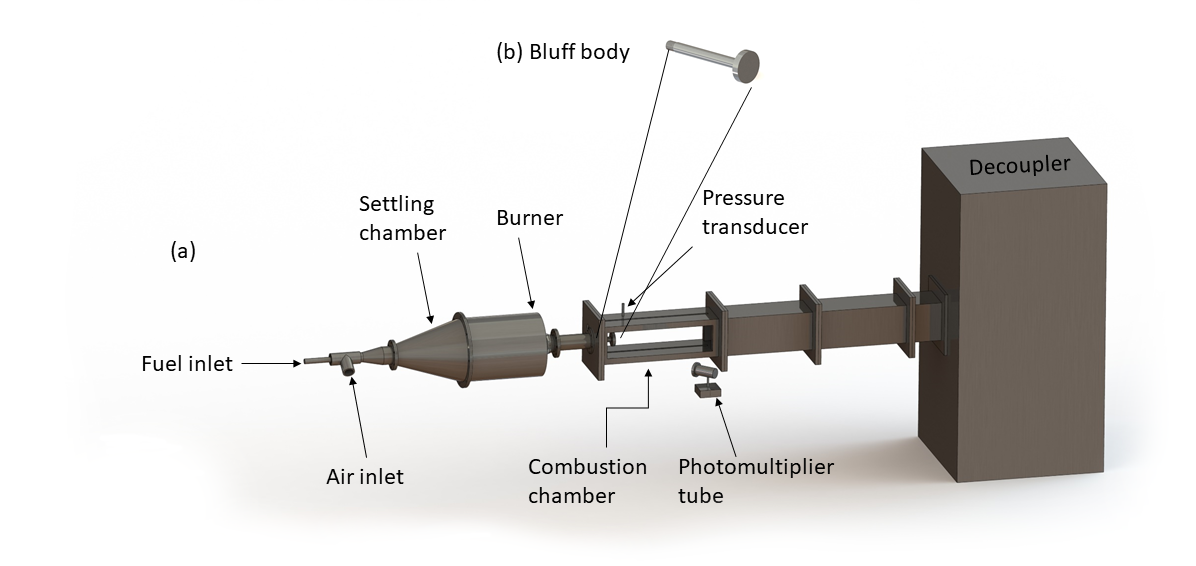}
    \caption{(a) The schematic of the turbulent combustor. (b) The schematic of the bluff body used in the experiments to stabilize the flame.}
    \label{fig:combustor}
\end{figure}

We performed experiments in a laboratory-scale turbulent combustor with a bluff body as a flame stabilizer. A schematic of the bluff body combustor used here is shown in Fig. \ref{fig:combustor}. The experimental setup consists of three main parts: a settling chamber, followed by the burner, where the fuel is injected and mixed with air, leading into the combustion chamber. 

The compressed air enters the settling chamber through the inlet. Here, the flow fluctuations in air are minimized. A central shaft delivers fuel (liquefied petroleum gas comprising 60\% butane and 40\% propane) via four radial injection holes with diameters of 1.7 mm each into the burner. The fuel mixes with air in the burner and flows into the combustion chamber.

The combustion chamber is a square duct of length 1100 mm with a cross-section of 90 mm $\times$ 90 mm. The reactant mixture entering the combustion chamber through the burner goes through a backward-facing step. A spark plug attached to the backward-facing step ignites the reactant mixture here. A circular bluff body having a diameter of 47 mm and thickness of 10 mm is fixed at the end of a hollow shaft at a distance of 32 mm from the backward-facing step. A recirculating zone is formed downstream of the bluff body. The recirculation zone comprises vortices that enhance mixing of hot radicals and reactant mixture. The heat release due to the combustion of the fuel-air mixture in these vortices adds energy to the fluctuations in the acoustic field. The combustion chamber is followed by the decoupler, which is a rectangular chamber of size 1000 mm $\times$ 500 mm $\times$  500 mm and is used to attenuate the fluctuations from the compressor thus isolating the combustion chamber. We maintain ambient pressure at the exit of the duct. Two quartz glass windows of 90 mm $\times$  360 mm are attached to the combustion chamber to provide optical access.

During the emergence of order, the fundamental mode of the combustion chamber is excited with a frequency of 160 Hz ($f=c/4L$, where $L$ is the length of the combustor and $c$ is the speed of sound in the combustion chamber). The mass flow rates of fuel ($\dot{m}_f$) and air ($\dot{m}_a$) are expressed in grams per second (g/s) and are controlled by mass flow controllers manufactured by Alicat Scientific, the MCR 2000 SLPM series for air and the MCR 100 SLPM series for fuel. The measurement uncertainty of the mass flow controllers is ± (0.8\% of the reading + 0.2\% of the full scale). Therefore, the uncertainty in the equivalence ratio is $\pm0.02$. For our experiments, $\dot{m}_a$ is varied from 15.3 g/s to 29.5 g/s and and $\dot{m}_f$ is kept constant at 1.75 g/s. To measure the acoustic pressure oscillations, the piezoelectric transducer (PCB103B02 with a sensitivity of 230.6 mV/kPa and a measurement uncertainty of  ±0.15  Pa) is mounted on the combustor wall at a distance of 40 mm downstream of the backward-facing step of the combustor using a T-joint mount. The global heat release rate fluctuations are measured using a photomultiplier tube (PMT) module. 

The Reynolds number in our study is varied from $3.9-6.8\times 10^4$. The Reynolds number is defined as $Re=(\rho \Bar{v} D)/\mu$, where $\Bar{v}$ represents the bulk velocity of the reactant mixture exiting the burner, $D$ is the diameter of the bluff body, $\mu$ is the dynamic viscosity of the mixture\cite{wilke1950viscosity} and $\rho$ is the density of the reactant mixture. The bulk velocity is defined as $\Bar{v}=Q/A$, where $Q$ is the flow rate and $A$ is the cross sectional area. The $Re$ calculation has a maximum uncertainty of $\pm 308$ based on the uncertainty of the mass flow controllers. In this study, we discuss results obtained from different equivalence ratios ($\phi$). The values of $\phi$ are calculated as :
\begin{equation}\label{equivalence ratio}
    \phi=\frac{(\dot{m}_f/\dot{m}_a)_{actual}}{(\dot{m}_f/\dot{m}_a)_{stoichiometric}}
\end{equation} 
The $\phi$ measured in this experiment is reduced from 1 to 0.52. The uncertainty present in the equivalence ratio is $\pm$ 0.02 based on the uncertainties of the mass flow controllers. To maintain repeatability of the experiments, we keep the acoustic damping within $\pm10\%$ of the exponential decay rate of the system at cold flow conditions.

\section{Forced Stuart-Landau oscillator model for the turbulent thermoacoustic system}\label{Methods}

We model the features of two global oscillatory variables of a turbulent thermoacoustic system, namely, the acoustic pressure ($p'$) and the global heat release rate fluctuations ($\dot{q}'$) during a dynamical transition from chaos to self-sustained limit cycle via intermittency. Each oscillatory variable is modeled as a Stuart-Landau oscillator \cite{garcia2012complex} driven by a forcing term representing the effect of inter-subsystem interactions.
\begin{equation}\label{eqn:SLO Noise}
    \dot{W_p}=\mu_p(\sigma+i \omega)W_p-(g_r+i g_i)|W_p|^2W_p+F_p
\end{equation}
\begin{equation}\label{eqn:SLO MMPP}
    \dot{W_q}=\mu_q(\sigma+i \omega)W_q-(g_r+i g_i)|W_q|^2W_q+F_q
\end{equation}
The forced Stuart-Landau oscillators in Eq. (\ref{eqn:SLO Noise}) and (\ref{eqn:SLO MMPP}) represent $p'$ and $\dot{q}'$, respectively. Here, $W_p$ and $W_q$ are complex variables, the parameters $\mu_p$ and $\mu_q$ are the control parameters, $\sigma$ is the amplitude growth rate, $\omega$ is the angular frequency, $g_r$ is the Landau constant, $g_i$ is the phase growth rate, and $F_p$ and $F_q$ are the forcing terms representing the holistic effect of inter-subsystem interactions on $p'$ and $\dot{q}'$, respectively. We use $\mu_p=\mu_q$, since both $p'$ and $\dot{q}'$ undergo a transition simultaneously when the flow control parameter is varied in a turbulent thermoacoustic system in the experiments. 

Note, a Stuart-Landau oscillator undergoes a supercritical Hopf bifurcation at $\mu=0$ for $g_r>0$ and is subcritical Hopf bifurcation at $\mu=0$ for $g_r<0$\cite{garcia2012complex}. In this study, we use the supercritical form of the Stuart-Landau oscillator with the parameters $\sigma=1$, $\omega=1/\mu$, $g_r=1$, $g_i=0$. We integrate the differential equation numerically using a Runge-Kutta method\cite{richardson2009stochastic} of fourth order with a timestep of $10^{-4}$ for 1000 non-dimensional time steps ($\tau$) with an initial condition for $W$ of the order $0.001+0.001i$ (we maintain the same small initial condition for all simulations).
\begin{equation}\label{eqn:noise pressure}
    p'=A_p \{\text{Re}[W_p(\tau\alpha)]\}
\end{equation}
 \begin{equation}\label{eqn: mmpp model}
     \dot{q}'=A_q\{\text{Re}[W_q(\tau\alpha)]-<\text{Re}[W_q(\tau\alpha)]>\}
 \end{equation}
 The $p'$ and $\dot{q}'$ time series are obtained by re-scaling the time steps $\tau$ in the model with $\alpha=10^{-3}~\text{s}$ in Eq. (\ref{eqn:noise pressure}) and (\ref{eqn: mmpp model}), to obtain the timescale $t=\alpha \tau$ similar to that in the experiments. In order to compare with signals from the experiments, the time series of both variables are normalized by the maximum amplitudes of the respective time series ($A_p,A_q$), and hence are non-dimensional.

\raggedbottom
 \subsection{Modeling the effect of inter-subsystem interactions on acoustic pressure fluctuations}\label{p'methods}
 
In experiments, both the flow and flame are turbulent in nature, and turbulence introduces multiple timescales in the acoustic pressure fluctuations ($p'$). During the occurrence of chaotic dynamics (combustion noise), $p'$ exhibits a wide power spectrum with a wide range of frequencies induced by the underlying turbulence. Also, $p'$ exhibits a broad multifractal spectrum\cite{nair2014multifractality}. During the emergence of order (transition to thermoacoustic instability), despite the presence of turbulence, we see the emergence of a single dominant peak in the power spectrum of $p'$ and the loss of multifractality \cite{nair2014intermittency}. This loss of multifractality quantified by the scaling of the Hurst exponent with variation in the amplitude of the $p'$ oscillations exhibits a universal scaling relation\cite{pavithran2020universality} with a power law exponent close to $-2$.

We use a Stuart-Landau oscillator driven by additive white Gaussian noise to model the influence of the underlying turbulent flow and turbulent flame dynamics on $p'$. In Eq. (\ref{eqn:noise pressure}), we use $F_p=\eta\xi$, where $\xi$ denotes the term generated by Wiener process\cite{levy2020wiener} and $\eta$ denotes the intensity; we use $\eta=0.01$ (effect of using different values of $\eta$ is discussed in Appendix \ref{App_Hurst_diffnoise}).


\subsection{Modeling the effect of inter-subsystem interactions on global heat release rate fluctuations}\label{q'methods}

The heat release rate fluctuations in a turbulent thermoacoustic system depend on the flame front oscillations. The flame front dynamics is influenced by the multi-scale vortical structures shed in the flow that distort the flame front. Interesting multifractal flame front structures are formed during the state of thermoacoustic instability \cite{raghunathan2020multifractal}. Vortices form and shed in recirculation zones and govern the mixing and convection of the unburnt fuel-air mixture in the flow. Fuel-air mixture contained in the vortical structures undergoes combustion after sufficient mixing within the core of the vortices, leading to a sudden release of large amounts of heat. Thus, we observe spikes (kicks) in the heat release rate fluctuations ($\dot{q}'$). Small vortices are shed aperiodically during chaotic dynamics, whereas large vortices are shed almost periodically during ordered dynamics \cite{George_Unni_Raghunathan_Sujith_2018}. Clearly, the stochasticity in the vortex shedding process disappears and vortex shedding becomes more and more periodic during the transition from chaos to order in the acoustic pressure dynamics.

We model the `kicks' or `spikes' observed in $\dot{q}'$ as events of an arrival process. Generally, an arrival process is represented by a time-varying Poisson process, where the average arrival rate varies with time. In a Poisson process, given the average arrival rate ($\lambda$), the probability of $k$ number of events occurring in a given time interval $P(k)$ is calculated using Eq. (\ref{eq_poisson}).
 \begin{equation}\label{eq_poisson}
     P(k)=\frac{e^{-\lambda} \lambda^k}{k!}
 \end{equation}   
We quantify the arrival rates during each dynamical state from experiment data. Note that, owing to changes in the vortex shedding process in the flow, the stochasticity of the arrivals also changes as order emerges in the system. We calculate short time average arrival rates during each dynamical state (see Appendix \ref{appA_MMPP_expt}) and show that multiple arrival rates exist during the state of chaotic dynamics due to the aperiodicity in the vortex shedding process during this state. However, as order emerges, the number of such average arrival rates reduces and the arrival of kicks becomes periodic during the state of thermoacoustic instability. 
 
Next, we model the time-variations in the arrival rates as Markov modulations in the arrival rates of the Poisson process. Such a process is referred to as a Markov modulated Poisson process (MMPP)\cite{fischer1993markov}. The average arrival rates are treated as states of a Markov chain and the switching between any two states is governed by a transition probability matrix. In our model, we use the average arrival rates ($\Lambda=\{\lambda_i\}$) given in Eq. (\ref{eqn: MMPP lambda}) and the transition probability matrix ($\mathbf{T}$) given in Eq. (\ref{eqn :tpm}). Note, $T_{ij}$ denotes the probability of transition between arrival rates $\lambda_i$ and $\lambda_j$. 
 \begin{equation}\label{eqn: MMPP lambda}
     \Lambda=\{50,40,35,25,20\}
  \end{equation}
\begin{equation}\label{eqn :tpm}
    \mathbf{T}=\begin{bmatrix}
    g(\mu) & h(\mu)& h(\mu)& h(\mu)& h(\mu)\\
    g(\mu) & h(\mu)& h(\mu)& h(\mu)& h(\mu)\\
    g(\mu) & h(\mu)& h(\mu)& h(\mu)& h(\mu)\\
    g(\mu) & h(\mu)& h(\mu)& h(\mu)& h(\mu)\\
    g(\mu) & h(\mu)& h(\mu)& h(\mu)& h(\mu)
    \end{bmatrix}
\end{equation}
where $g(\mu)=1-e^{-(10\mu+5)}$, and $h(\mu)=(e^{-(10\mu+5)})/4$. The forcing term in Eq. (\ref{eqn:SLO MMPP}), can then be written as $F_q=\beta M(\Lambda,\mathbf{T})$ where $M$ is an MMPP process. We use $\beta=0.035$. Note that, the number of and range of values of arrival rates in $\Lambda$, and the probability functions that are dependent on the control parameter ($\mu$) inbuilt in $\mathbf{T}$ are inspired from the analysis of the experimental data presented in Appendix \ref{appA_MMPP_expt}. To model the change in the distribution of the arrival rates across different dynamical states, we quantify the transition probability functions from one arrival rate to another through experimental data in Appendix \ref{appA_MMPP_expt}. We show that, during chaotic dynamics, transitions between various average arrival rates are significant. However, as order emerges, only one of the arrival rates is dominant and hence the transition probabilities become zero from all other states. This understanding is encoded in $\mathbf{T}$ via functions $g(\mu)$ and $h(\mu)$.
\section{Results and discussion}\label{Results}

Here, we describe the results obtained from our model and compare the characteristic features of the time series of acoustic pressure ($p'$) and heat release rate ($\dot{q}'$) oscillations obtained from the model (Eq. (\ref{eqn:noise pressure}) and (\ref{eqn: mmpp model})) with those observed in the experiments. 

\subsection{Noise driven oscillator models the acoustic pressure oscillations in a turbulent reacting flow}\label{sec_noise_result}

\begin{figure}[h!]
    \includegraphics[width=\textwidth]{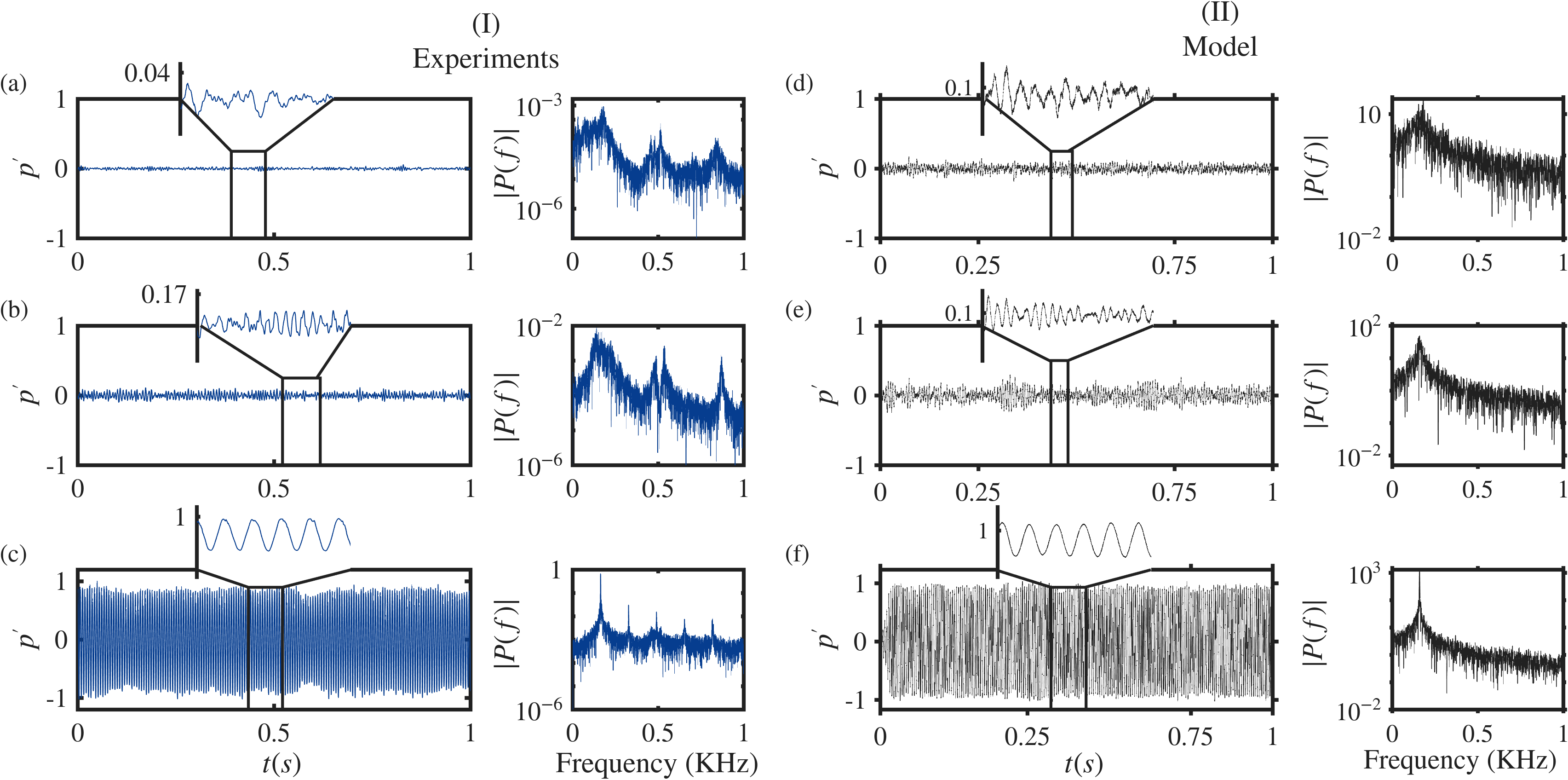}
    \caption{ Comparison of the time series of acoustic pressure fluctuation ($p'$) obtained from the experiments, normalized with the maximum amplitude of oscillation in the limit cycle regime (column (I)), from the model as per Eq. (\ref{eqn:noise pressure}) (column (II)), and the corresponding power spectrum. For the experiments, the mass flow rate of air is increased and thus the equivalence ratio $\phi$ decreases from (a) $\phi=1$, to (b) $\phi=0.76$,  to (c) $\phi=0.56$, respectively. In the model, the parameter $\mu$ is increased from (d) to (f) as $-0.2$, $-0.05$, and $0.3$, respectively, for a constant noise intensity of $\eta=0.01$. The time series of $p’$ obtained from the model replicate the characteristic features of the time series of $p’$ obtained from the experiments.}
 \label{fig:noiseffts}
\end{figure}

Figure \ref{fig:noiseffts} shows the time series of the acoustic pressure fluctuations ($p'$) obtained from the experiments (column-I) and from the model (column-II) by solving Eq. (\ref{eqn:noise pressure}) given the forcing term $F_p$ for different values of $\mu$ (Sec. \ref{p'methods}).
As the flow control parameter (the equivalence ratio $\phi$) is varied in the experiments, we observe a transition from chaos (see Fig. \ref{fig:noiseffts}(a)) to intermittency (Fig. \ref{fig:noiseffts}b) to self-sustained periodic oscillations (Fig. \ref{fig:noiseffts}c). Using the model, we replicate the characteristic features of the time series during each of these dynamical states as evident in Fig. \ref{fig:noiseffts}-II. Prior to, but far from the Hopf bifurcation point in the model, at $\mu=-0.2$ (see Fig. \ref{fig:noiseffts}(d)), we obtain a signal with low-amplitude aperiodic fluctuations similar to the state in Fig. \ref{fig:noiseffts}(a). The power spectrum of $p'$ obtained from the model exhibits a broadband spectrum of frequencies similar to that in experiments. Close to but prior to the Hopf point, at $\mu=-0.05$ (Fig. \ref{fig:noiseffts}(e)), we get a time series with intermittent bursts of periodic oscillations amidst low-amplitude aperiodic fluctuations. Finally, for positive values of $\mu$ far from the Hopf point, such as $\mu=0.3$ in Fig. \ref{fig:noiseffts}(f), we observe that the Stuart Landau oscillator representing $p'$ exhibits self-sustained limit cycle oscillations despite the presence of noise. The effect of the noise is evident in the amplitude modulation of the limit cycle as seen in Fig. \ref{fig:noiseffts}(f). Such amplitude modulated periodic signal is a typical feature of acoustic pressure signals in practical combustion systems \cite{kasthuri2019bursting, tandon2020bursting, pawar2016intermittencySPRAY}. Further, the power spectrum of $p'$ exhibits a sharp peak during this state corresponding to the frequency of the limit cycle oscillations.

\begin{figure}[h!]
    \centering
    \includegraphics[width=\textwidth]{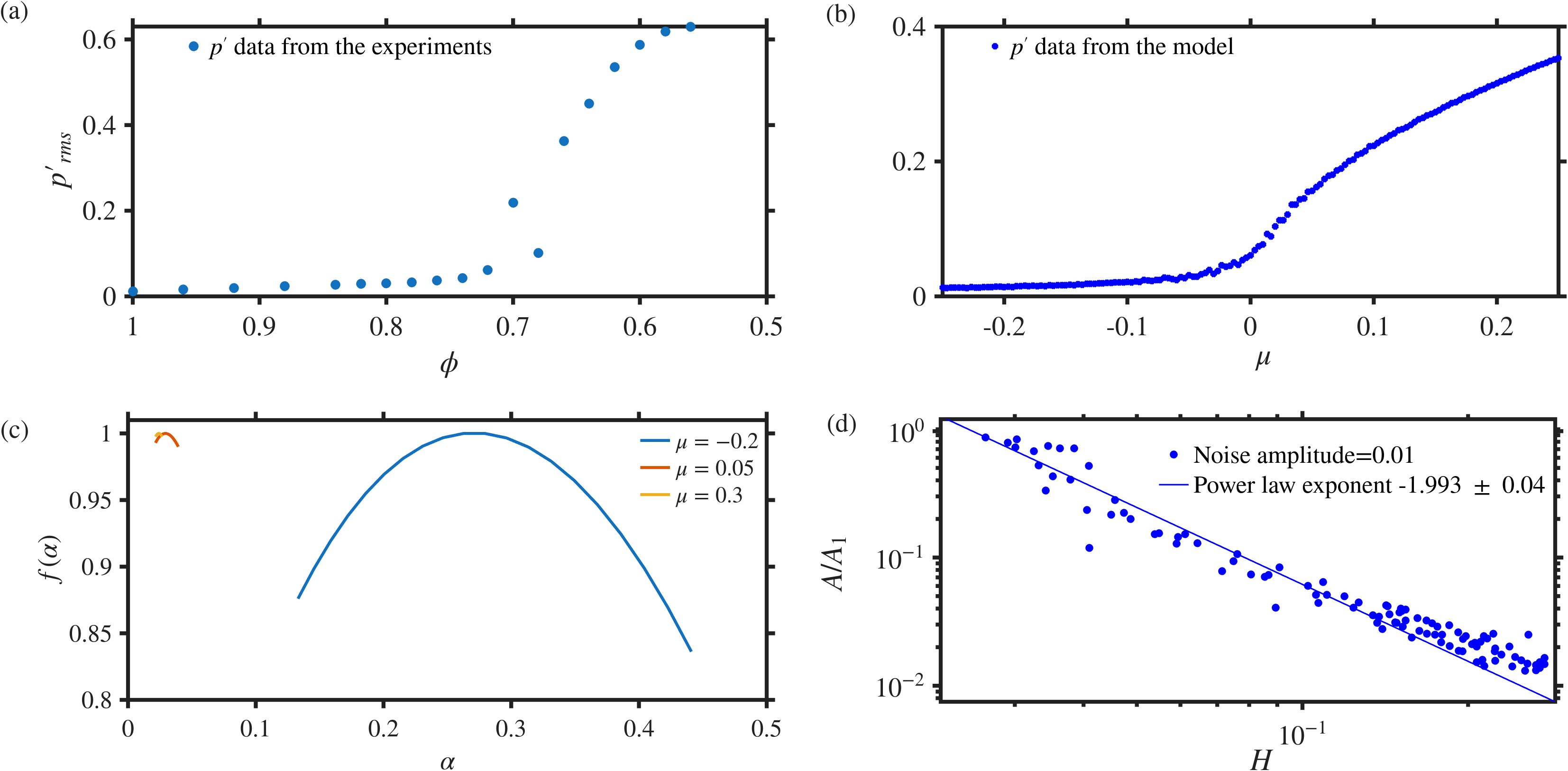}
    \caption{The bifurcation diagram for the normalized acoustic pressure fluctuations ($p’$) with respect to (a) flow control parameter ($\phi$) in the experiments and (b) the control parameter ($\mu$) for the model. A smooth bifurcation occurs at $\phi' = 0.74$ in experiments and at $\mu = 0$ in the model. (c) The multifractal spectrum of $p'$ signals obtained from the model at $\mu=-0.2,~0.05,\text{and}~0.3$. (d) Variation of the spectral amplitude ($A$) normalized with the spectral amplitude at the onset of thermoacoustic instability ($A_1$)  of the dominant frequency of $p'$ with the Hurst exponent ($H$) when $\mu \in [-0.25,0.05]$ in the model, exhibits a power law scaling with the exponent $-1.993\pm 0.04$ (95\% confidence). The multifractal spectrum shrinks as periodicity emerges and the Hurst exponent exhibits the universal scaling law observed from experiments in turbulent thermo-fluid and fluid systems \cite{pavithran2020universality}.}
 \label{fig: bifnoise}
\end{figure}

To study the bifurcation of the dynamics, we plot the variation of the root mean square of the acoustic pressure fluctuations ($p_{rms}’$) with a quasi-static change in the flow control parameter ($\phi$) for data from experiments (Fig. \ref{fig: bifnoise}(a)) and with respect to $\mu$ for the model (Fig. \ref{fig: bifnoise}(b)). The bifurcation diagram obtained from the model exhibits a continuous transition with a smooth increase in $p'_{rms}$ around the bifurcation point. Notice the small increase in $p'_{rms}$ even prior to $\mu=0$, emulating the bifurcation diagram from experiments owing to the occurrence of intermittency. 

Figure \ref{fig: bifnoise}(c) shows the multifractal spectra (refer Appendix \ref{app_Estimation_hurst} for detailed explanation) obtained for different dynamical states shown in Fig. \ref{fig:noiseffts}-II. Clearly, the low-amplitude aperiodic fluctuations in $p'$ obtained from the model exhibit a broad multifractal spectrum that shrinks during the state of intermittency and then condenses to a single point during limit cycle oscillations. Such loss of multifractality in the signals obtained from the model is an intriguing signature of the experiment data that was discovered by \textcite{nair2014multifractality}. Next, we use the Hurst exponent calculated from the multifractal spectrum to characterize the persistence of a signal. A value of the Hurst exponent ($H$) close to $0.5$ implies white noise characteristics, where $H<0.5$ implies the signal exhibits anti-persistence and $H>0.5$ implies the signal exhibits persistence. Persistent signals are positively correlated, showing a tendency for increases (or decreases) to be followed by further increases (or decreases), while anti-persistent signals are negatively correlated, tending to reverse direction such that increases are more likely to be followed by decreases, and vice versa\cite{sujith2021thermoacoustic}. As we increase $\mu$ in the range $[-0.25,0.05]$, we find that $H$ decreases following an approximately inverse square power-law scaling with the spectral amplitude of the dominant frequency of $p'$ signals obtained from the model. This power-law scaling replicates the universal scaling law reported for experiments in turbulent thermoacoustic, aero-acoustic and aeroelastic systems\cite{pavithran2020universality}. The method used in this study for calculating the Hurst exponent is described in Appendix \ref{app_Estimation_hurst} and the optimal choice of the noise intensity used in our model for replicating such a scaling relation is discussed in Appendix \ref{App_Hurst_diffnoise}.

Note that our model replicates all the characteristic features and the quantitative scaling exhibited by $p'$, that arises in experiments owing to the highly complex flow-fluid-acoustic interactions in the system. Evidently, the overall influence of the spatio-temporal web of inter-subsystem interactions in the system on the dynamics of $p'$ are well captured by our model, a Stuart Landau oscillator forced by the Wiener process at each step of integration. It is important to emphasize that the previous approaches to replicating such scaling law \cite{weng2023synchronization, garcia2012complex} are also robust and indeed represent a reduced-order model of the complex spatio-temporal dynamics by accounting for coupling across several variables and subsystems. Here, our objective is to model the features of a global oscillating variable as an independent nonlinearity coupled to the underlying web of interactions across subsystems. Thus, our model provides a novel reduced-order representation of $p'$ dynamics without introducing coupled oscillator systems as done in previous approaches, and yet captures the main qualitative and quantitative features of the dynamics. 

\subsection{MMPP driven Stuart-Landau oscillator models heat release rate fluctuations in a turbulent reacting flow} \label{sec_MMPP_result}

\begin{figure}[h!]
    \centering
    \includegraphics[width=\linewidth]{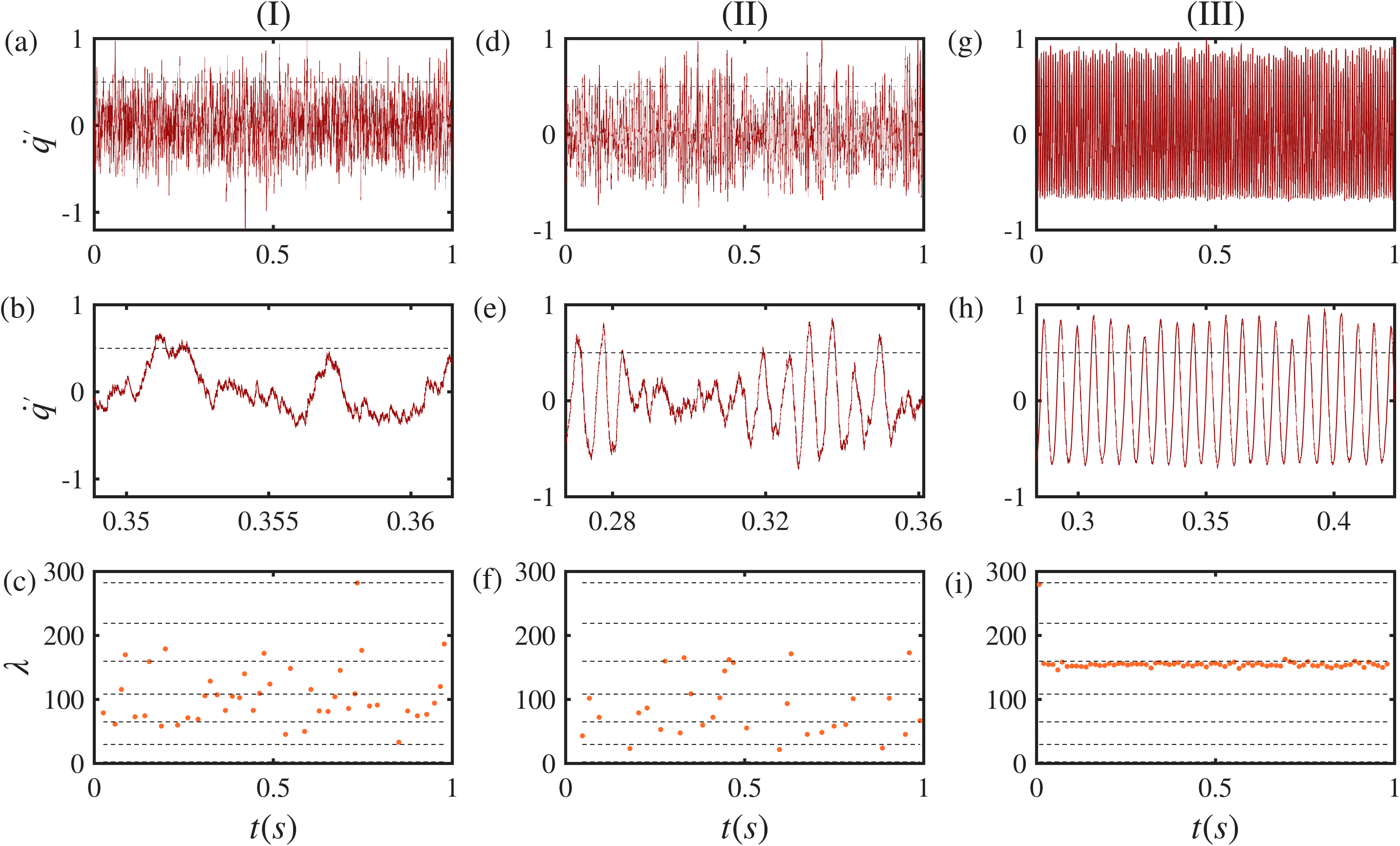}
    \caption{The time series of ((a),(d) and (g)) $\dot{q}’$ obtained from the experiments as the turbulent thermoacoustic system transitions from (I) low-amplitude aperiodic fluctuations to (III) high amplitude periodic fluctuations through (II) intermittency. The zoomed-in time series ((b),(e), and (h)) for $\dot{q}’$ illustrate the calculation of average arrival rates using a threshold and $k=3$ in Eq. (\ref{avg arrival rate}), therefore yielding an average arrival rate of $r=3/\Delta \tau$.  The average arrival rates ((c),(f), and (i)) binned using Lambda algorithm\cite{heyman2003modeling} plotted against time.}
    \label{fig:kickavgarrival}
\end{figure}

In this section, we use MMPP driven Stuart Landau oscillator (Eq. (\ref{eqn: mmpp model}), Sec. \ref{q'methods}) and replicate the characteristic features of the global heat release rate fluctuations ($\dot{q}'$) in a turbulent thermoacoustic system.

\begin{figure}[h!]
    \centering
    \includegraphics[width=\textwidth]{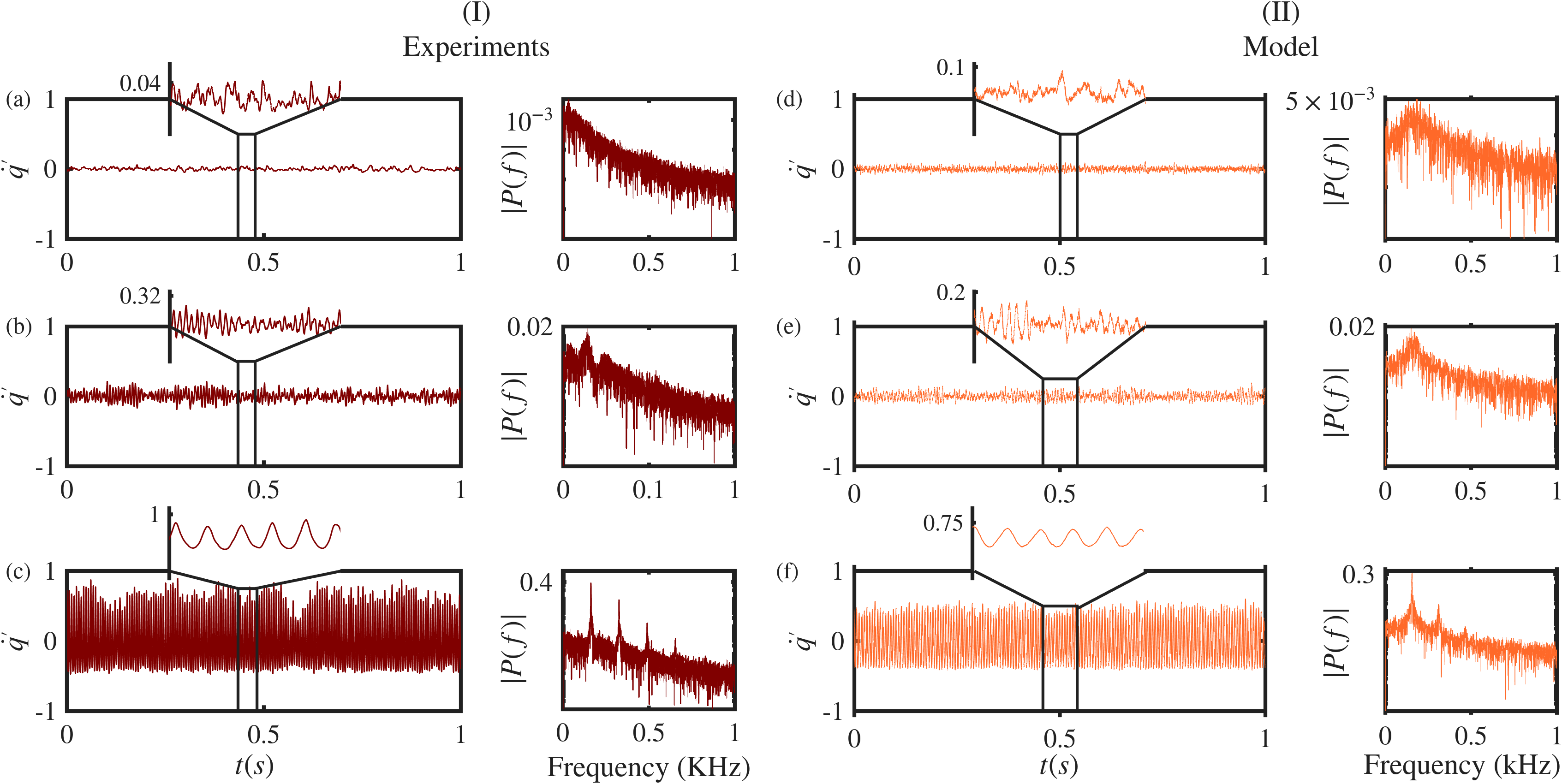}
    \caption{Comparison of the heat release rate fluctuation ($\dot{q}'$) time series obtained from the experiments, normalized with the maximum amplitude of oscillation in the limit cycle regime  (column (I)), from the model as per the Eq. (\ref{eqn: mmpp model}) (column (II)), and their power spectrum. For the experiments, the flow control parameter $\phi$ from (a) to (c) as 1, 0.76, and 0.56 respectively. In the model, the parameter $\mu$ is increased from (d) to (f) as -0.2, -0.002, and 0.4, respectively for a constant kick amplitude of $\beta=0.035$. }
 \label{fig: mmppfft}
\end{figure}

Column 1 in Fig. \ref{fig: mmppfft} shows the global heat release rate fluctuation ($\dot{q}’$) obtained from the experiments in the turbulent thermoacoustic combustor as the flow control parameter ($\phi$) is varied. 
We identify `kicks' in a time series of $\dot{q}’$ as local maxima above a set threshold (refer Appendix \ref{appA_MMPP_expt}). The kicks in $\dot{q}’$ are of low amplitude and are distributed irregularly during chaotic $p'$ dynamics (Fig. \ref{fig:kickavgarrival}(b)). As order emerges, we find that kicks occur almost periodically for short epochs during the state of intermittency (Fig. \ref{fig:kickavgarrival}(f)). Finally, during the ordered dynamics in $p'$, we observe order in $\dot{q}’$ in the form of high-amplitude periodic kicks (Fig. \ref{fig:kickavgarrival}(i)). The average arrival rate distribution obtained using our model is similar to that observed in experiments and is discussed in Appendix \ref{appA_MMPP_expt}.

\begin{figure}[h!]
    \centering
    \includegraphics[width=\textwidth]{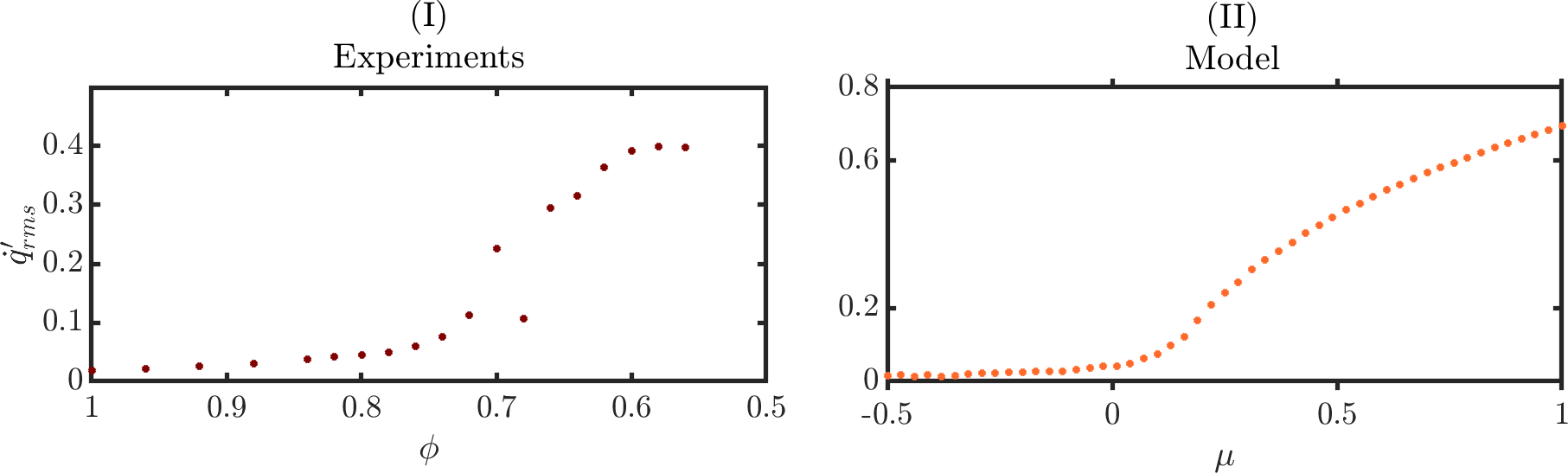}
    \caption{The bifurcation diagram for the global heat release rate fluctuations ($\dot{q}’$) with respect to (a) flow control parameter ($\phi’$) in the experiments and (b) the control parameter ($\mu$) for the model given by the Eq. (\ref{eqn: mmpp model}).}
 \label{fig: mmppbif}
\end{figure}

The time series of heat release rate fluctuation during various dynamical states obtained from the experiments (column-I) and the model (column-II) is shown in Fig. \ref{fig: mmppfft}, along with the corresponding power spectra. The time series for $\dot{q}’$ generated via this model exhibits low-amplitude irregular kicks at negative values of $\mu$, similar to the experiments. Upon increasing the value of $\mu$, the amplitude of the kicks increases, and the arrival of kicks becomes relatively periodic for short durations. Finally, for $\mu>0$, we observe almost periodic arrivals of kicks from the model and the $\dot{q}’$ signal is periodic.

We observe, that the power spectrum of $\dot{q}’$ has a broad peak for high values of $\phi$ in experiments (chaotic state). However, as we lower the value of $\phi$, order emerges, and the turbulent thermoacoustic system undergoes spectral condensation\cite{pavithran2020universality}, leading to a single dominant peak in the power spectrum. We find a similar evolution of the power spectrum of the $\dot{q}’$ signals obtained from the model (Fig. \ref{fig: mmppfft} column-II).

Finally, we plot the bifurcation diagrams (Fig. \ref{fig: mmppbif}) to compare the transition of $\dot{q}’$ dynamics in experiments and the model. The bifurcation diagrams (Fig. \ref{fig: mmppbif}) exhibited the variation of the root mean square of the $\dot{q}_{rms}’$ with a quasi-static change in $\mu$ for the model and the flow control parameter $\phi$ as varied from 1 to 0.56 in the experiments. When $\phi$ is decreased, the root mean square value of the global heat release rate fluctuations ($\dot{q}_{rms}’$) exhibits a continuous and smooth transition in experiments and model.

Evidently, the model is able to replicate the dynamics of the global heat release rate fluctuations. Thus, the underlying non-linearity in the flame dynamics subsystem is captured by the third-order non-linearity in the Stuart-Landau oscillator. Most importantly, we model the change in stochasticity of the spikes in the heat release rate signal. These spikes in the heat release rate are attributed to vortex shedding and the subsequent combustion of the fuel-air mixture contained in the vortices. We use a transition probability matrix that varies with the control parameter $\mu$, to encode the change in stochasticity of arrival of kicks (and hence, the vortices). The evolution of the transition probability matrix ensures that the periodicity in the arrival of kicks increases as $\mu$ increases, and hence we find that order emerges in the system. The distribution of kicks obtained using MMPP is qualitatively similar to the distribution of vortex shedding times (Fig. \ref{fig:modelarr}).

 
\section{Conclusion}\label{Conclusion}

In this study, we investigate the effect of complex inter-subsystem interactions in a turbulent thermoacoustic system through a conceptual reduced-order model. Using Stuart-Landau oscillators, we model the nonlinearity in the global oscillating variables of the turbulent thermoacoustic system, namely, the acoustic pressure and the global heat release rate oscillations. These oscillators are driven by a forcing term that is determined by analyzing the interactions between the various subsystems of the turbulent thermoacoustic system. Each of both Stuart-Landau oscillators have unique forcing terms to distinguish the holistic effects of inter-subsystem interactions in the system.

We describe the emergent global dynamics in the acoustic pressure signal ($p'$) using this oscillator driven by Gaussian white noise. The cubic nonlinearity of this oscillator captures the smooth bifurcation in $p'$ dynamics. Further, the forcing by noise emulates the multifractality during low-amplitude aperiodic dynamics in $p'$. Similar to experiments, the multifractality in the $p'$ signal obtained from the model is lost as order emerges and oscillator dynamics bifurcates to limit cycle oscillations. This loss of multifractality in the model occurs via the universal scaling relation obtained in practical turbulent fluids and thermo-fluid systems. The emergent dynamics in the global heat release rate signal delineates multiple intriguing features that we capture using a Stuart Landau oscillator driven by time-varying arrival process. We use a Markov-modulated Poisson process, where the arrival process represents the arrival of a fuel-laden vortex and subsequent combustion in a turbulent thermoacoustic system. We note that in experiments we vary only one control parameter; however, the periodicity and strength of the vortex shedding in flow and the interactions of the acoustic pressure with the flame change inadvertently. This way we capture such complexity systematically in our model, where we can vary only one control parameter. However, the arrival rates and transition probabilities of the Markov chain vary with the control parameter of the model. We are able to replicate well the bifurcation and characteristic features of the heat release rate signal as observed in experiments. 

Our reduced-order modeling can also be valuable in understanding other turbulent systems, such as turbulent aeroelastic and aero-acoustic systems. Moreover, such a reduced order model can be adapted for an effective reduced-order system identification in general complex turbulent flow systems. Further, the novelty of our approach lies in modeling the complexity of the system using an appropriate nonlinearity driven by an effective term representing the holistic effects of interactions in the system. Such an approach can find potential in modeling other complex systems as well.

\vspace{50pt}
{{\textbf{Data Availability:}} The experiment data presented in this work is the same as that published by \textcite{sudarsanan2024emergence}. A code for studying the dynamical evolution of Stuart Landau oscillation with MMPP forcing is available on 
\newline\textcolor{blue}{[https://github.com/ae20b011/ThermoacousticSLE.git]}
}

{{\textbf{Acknowledgment}}
The authors thank Mr. Sivakumar, Mr. Midhun, Ms. Anaswara, Mr. Tilag Raj, Mr. Anand, and Ms. Sudha for providing the experimental data. This research was supported by grants provided to R.I.S., namely, J. C. Bose Fellowship (JCB/2018/000034/SSC) and the IoE-IITM Research Initiatives (SP/22-23/ 1222/ CPETWOCTSHOC). S.T. acknowledges the support from Prime Minister's Research Fellowship, Govt. of India. Authors also thank Mr. Jayesh Dhadphale and Dr. Induja Pavithran for the fruitful discussions and valuable suggestions. }

{{\textbf{Author contributions}} Conception: S.T., R.I.S., Methodology and analysis: A.S., A.E.B., S.T., Interpretation: S.T., A.S., A.E.B., J.K., N.M., R.I.S. Visualization: A.S., Supervision: J.K., N.M., R.I.S., Writing original draft: A.S., S.T., Review and editing: R.I.S., A.E.B., J.K., N.M.}

\newpage
\appendix 
\section{Estimation of parameters of the MMPP}\label{appA_MMPP_expt}

We estimate the parameters for the MMPP from the time series of $\dot{q}'$ obtained from experiments. To estimate the average arrival rates, we set a threshold of 0.6 on the normalized $\dot{q}'$ time series and get the timestamps of `kicks' as the locations of the local maxima above this threshold, as shown in Fig. \ref{fig:kickavgarrival}. The threshold value was chosen to be 1.5 times the standard deviation of the normalized heat release rate fluctuation timeseries for thermoacoustic instability. Let the time taken for $k$ events to occur be $\Delta \tau$. The average arrival rate is given by Eq. (\ref{avg arrival rate}). 
\begin{equation}\label{avg arrival rate}
    r=\frac{k}{\Delta \tau}
\end{equation}
In order to determine the local arrival rates, we use $k=3$. Very high values of $k$ will lead to a loss in the number of arrival rates and hence induce a bias in the estimation of transition probabilities between arrival rates. The value  $k=3$ was chosen to meet these constraints following \textcite{barr1998straightforward}.

We use the binning method proposed by \textcite{heyman2003modeling}, where the highest and lowest values of $r$ are chosen, and a width parameter `$a=2$’ is chosen based on \cite{heyman2003modeling} to bin the values of $r$. The bin width determines the number and range of arrival rates. The average arrival rates are then calculated using the Eq. (\ref{lambda}).
\begin{equation}\label{lambda}
     \lambda_j = \left(\frac{(\sqrt{(a^2+4\times \lambda_{j+1})}-a)}{2}\right)^2
\end{equation}
For calculating the average arrival rate of the first bin, $\lambda_{j+1}$ is replaced by $\lambda_{max}$, the highest value of $r$. The upper limit of a bin for a particular average arrival rate $\lambda_j$ is $\lambda_j+a\sqrt{\lambda_j}$, and the lower limit of the bin is $\lambda_j-a\sqrt{\lambda_j}$.

The transition probability matrix is estimated from the binned average arrival rates as shown in Eq. (\ref{transition probability matrix})
\begin{equation}\label{transition probability matrix}
    P_{ij}=\frac{T_{ij}}{\sum_k^n T_{ik}}
\end{equation}
Here, $T_{ij}$ denotes the number of times the state transitions from $i$ to $j$, and $\sum_k^n T_{ik}$ denotes the total number of times the state transitions to $i$.

\begin{figure}[h!]
    \centering
    \includegraphics[width=0.95\linewidth]{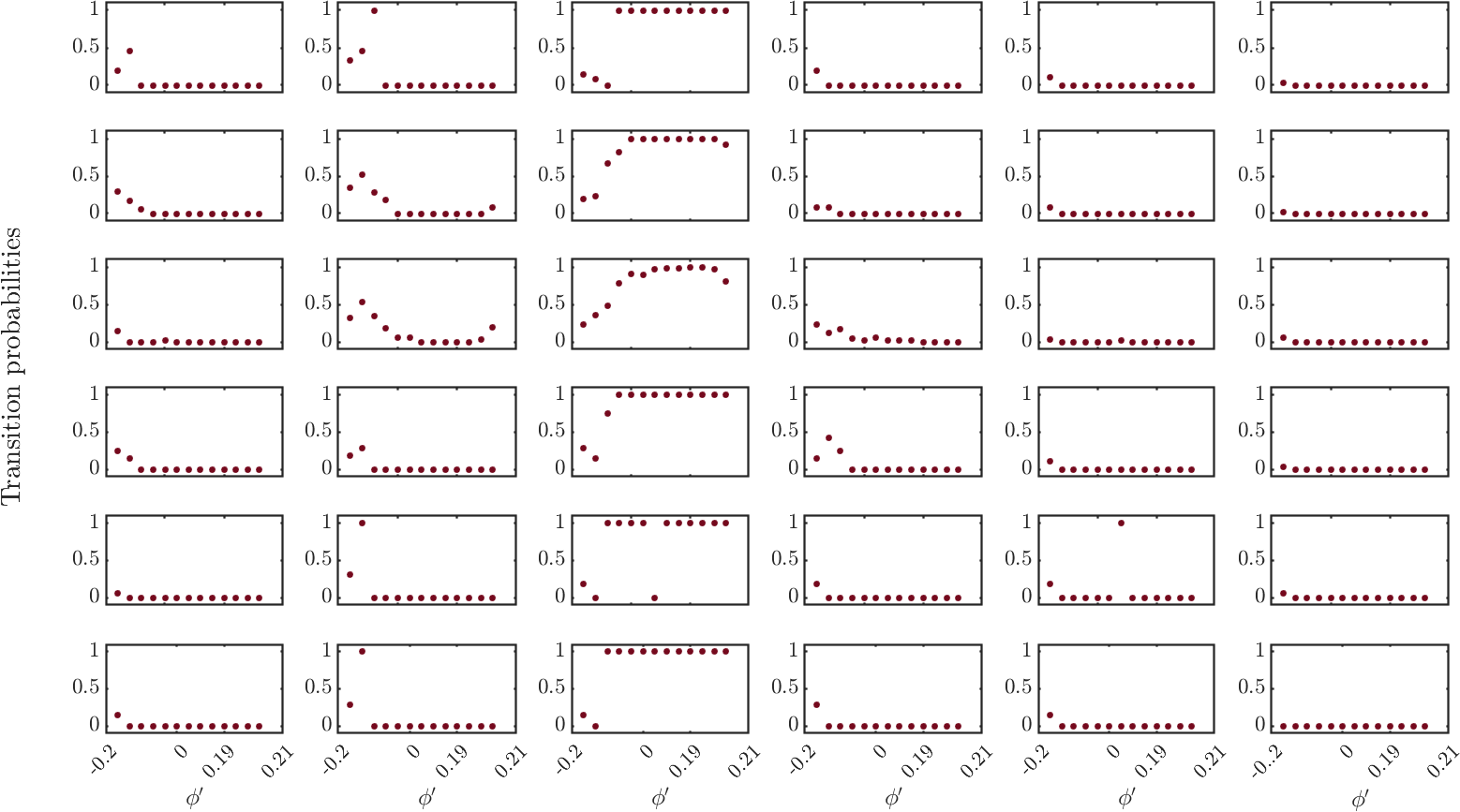}
    \caption{The variation of transition probabilities in the transition probability matrix with the flow control parameter $\phi’=\phi_{cr}-\phi$. Each plot represents the variation of an element of the transition probability matrix as the system transitions from low amplitude aperiodic fluctuations in $\dot{q}’$ to high amplitude periodic fluctuations.}
    \label{fig: tpmdist}
\end{figure}
\begin{figure}[h!]
    \centering
    \includegraphics[scale=0.55]{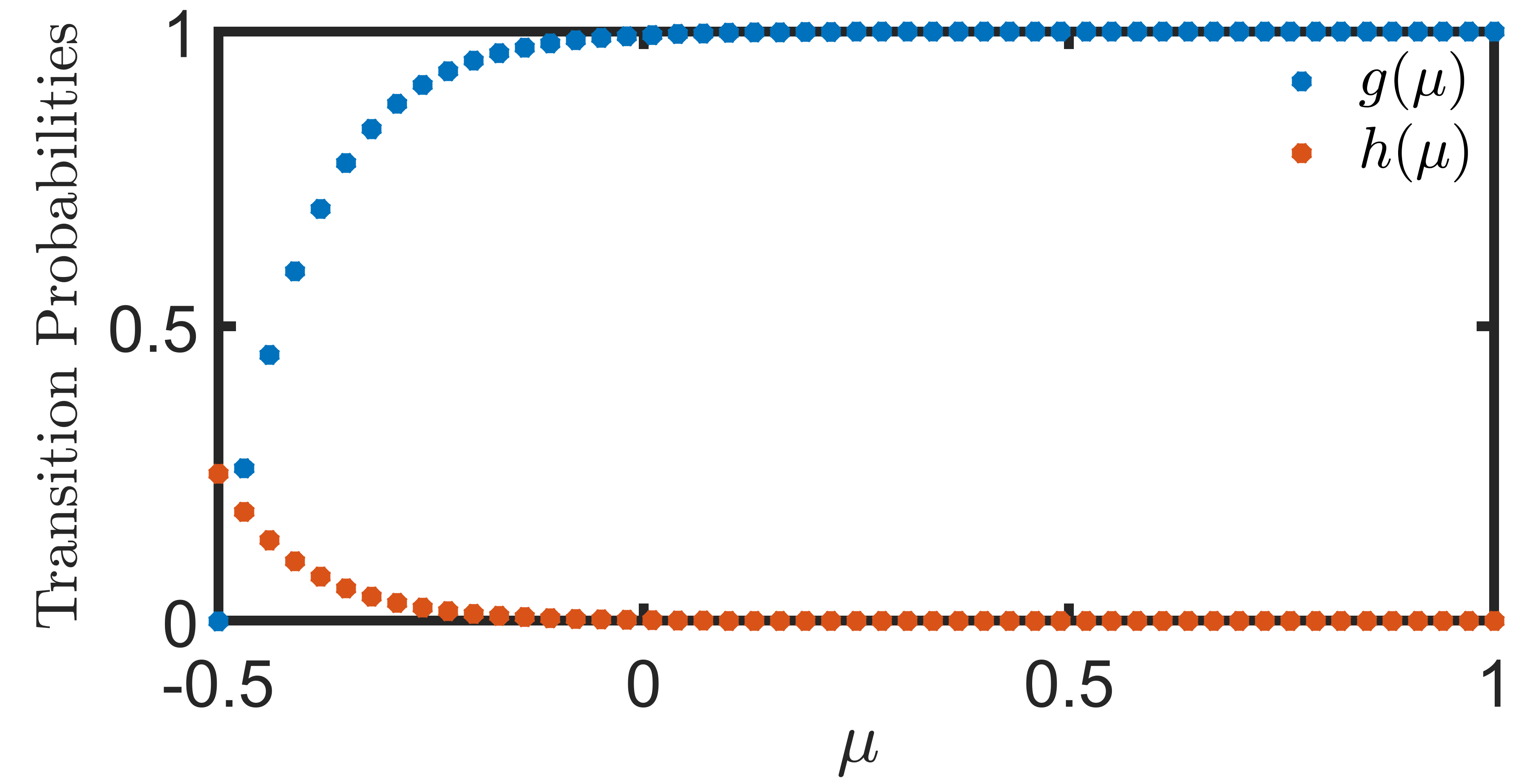}
    \caption{The variations of the functions $g(\mu)$ and $h(\mu)$ with the control parameter $\mu$, as used in Eq. (\ref{eqn :tpm}) .}
    \label{fig:variationf1f2}
\end{figure}

Figure \ref{fig: tpmdist} clearly shows that as the system becomes more and more periodic, the transition probabilities vary significantly for different arrival rates. For all the arrival rates except one, the transition probabilities go to zero as the system becomes periodic. This implies that initially, the switching between the average arrival rates occurs uniformly as the transition probabilities are uniformly distributed, but as the flow control parameter is increased, there is almost no switching between average arrival rates, and one average arrival rate dominates the others. Keeping in mind these observations, we arrived at the transition probability matrix shown in Eq. (\ref{eqn :tpm}), where $g(\mu)$ and $h(\mu)$ vary as shown in the Fig. \ref{fig:variationf1f2}.

\begin{figure}[h!]
    \centering
    \includegraphics[width=\linewidth]{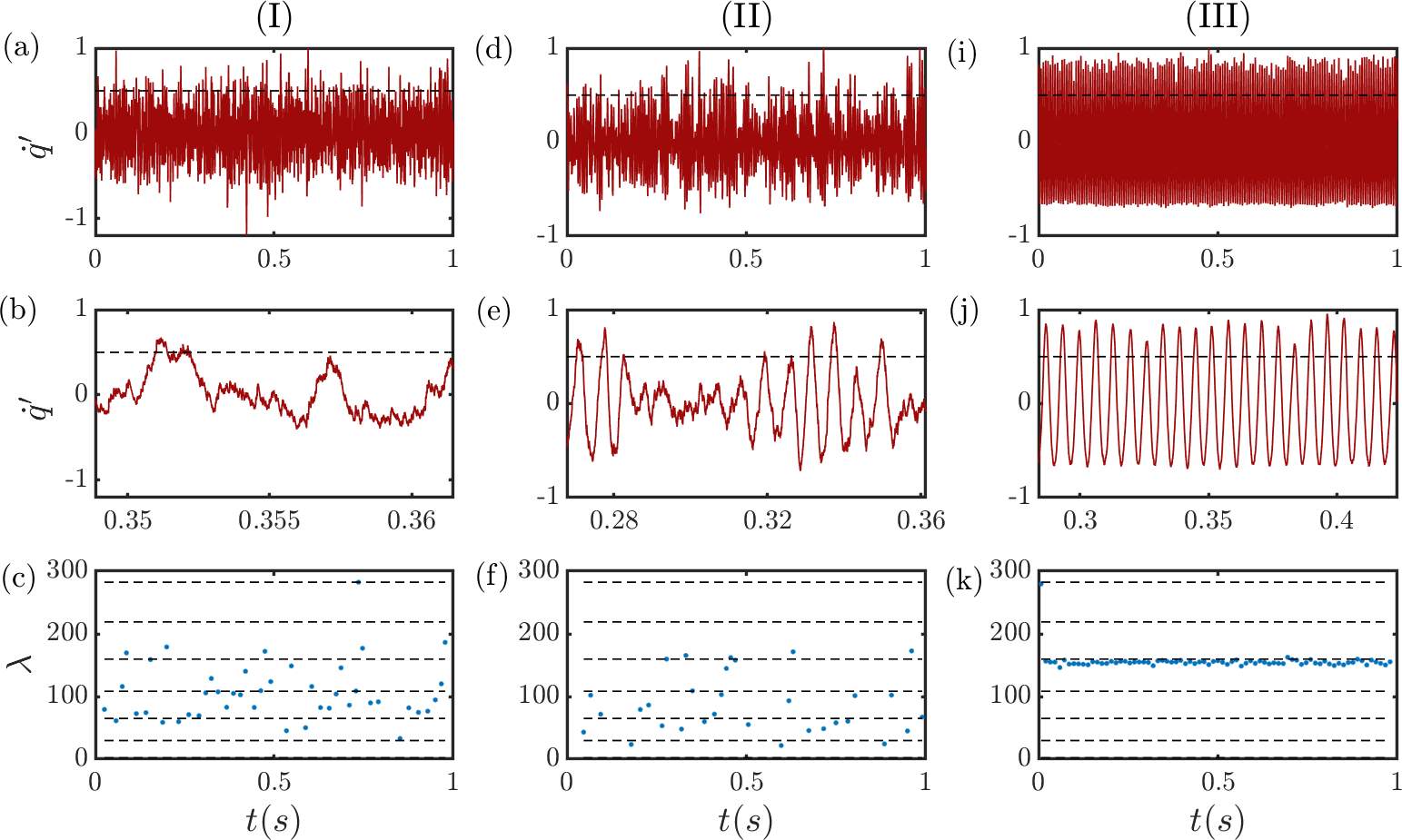}
    \caption{(a),(d) and (g) shown the time series of $\dot{q}’$ obtained from the model as $\mu$ is increased. The dynamics of the time series for $\dot{q}'$ transitions from (I) low-amplitude aperiodic fluctuations to (III) high-amplitude periodic fluctuations through (II) the state of intermittency. The zoomed-in time series is shown in (b),(e), and (h) and illustrate the calculation of average arrival rates using a threshold and $k=2$ in Eq. (\ref{avg arrival rate}). The average arrival rates plotted in (c),(f), and (i) are binned using Lambda algorithm\cite{heyman2003modeling} plotted against time. The change in stochasticity in vortex shedding observed during experiments is incorporated into the model through MMPP.}
    \label{fig:modelarr}
\end{figure}

The features of the arrival of kicks in the heat release rate signal obtained from our model are shown in Fig. \ref{fig:modelarr}. Clearly, the features of the average arrival rates obtained from the model are similar to that obtained from experiments as shown in Fig. \ref{fig:kickavgarrival}.

\section{Estimation of the Hurst exponent and multifractal spectrum}\label{app_Estimation_hurst}

Hurst exponent is a statistical measure for classifying a time series according to its long term memory\cite{qian2004hurst}\cite{nair2014multifractality}. The most popular algorithms for estimating the Hurst exponent are multifractal detrended fluctuation analysis (MFDFA)\cite{kantelhardt2002multifractal}, rescaled range analysis (R/S)\cite{hurst1951long}, wavelet transform\cite{carbone2004time}, and multifractal detrended moving average (MFDMA)\cite{gu2010detrending}. We use MFDFA method for estimating the Hurst exponent as it is robust to trends in the time series. The procedure to calculate the q-order Hurst exponent (q=2) using MFDFA as given by \textcite{ihlenmfdfa} and the estimation of multifractal spectrum given by \textcite{kantelhardt2002multifractal} are described below.

Given a time series $x(t)$ of length $N$, we define a mean subtracted cumulative deviation series $Y(k)$ as,
\begin{equation}
    Y(k)=\sum^k_{t=1}{[x_t-<x>]}, \hspace{1cm}k=1,2,....N
\end{equation}
Here $<x>$ is the mean of the time series. We divide the series $Y(k)$ into equal parts of length $w$ each. We then remove the local trends by subtracting the polynomial fit for each segment. We use a polynomial fit of order 1 in this study. The variance of the detrended series ($\Bar{Y_i}$) is given by Eq. (\ref{variancedetrended}).
\begin{equation}\label{variancedetrended}
    F^2(w,i)=\frac{1}{w}[\sum^w_{t=1}{(Y_i(t)-\Bar{Y_i})^2}]
\end{equation}
$F_w$ can then be determined as,
\begin{equation}
 F_w^2=[\frac{1}{N_w}\sum^{N_w}_{i=1}{F^2(w,i)}]^\frac{1}{2}  
\end{equation}
Here, $N_w$ is the number of segments of length $w$. We repeat the process for multiple values of the span $w$. The slope of the linear regime in the curve $F_w$ vs $w$ in a log-log plot gives the Hurst exponent (H).
\begin{equation}
    F(w) \propto w^{-H}
\end{equation}

\begin{figure}[h!]
    \centering
    \includegraphics[scale=0.4]{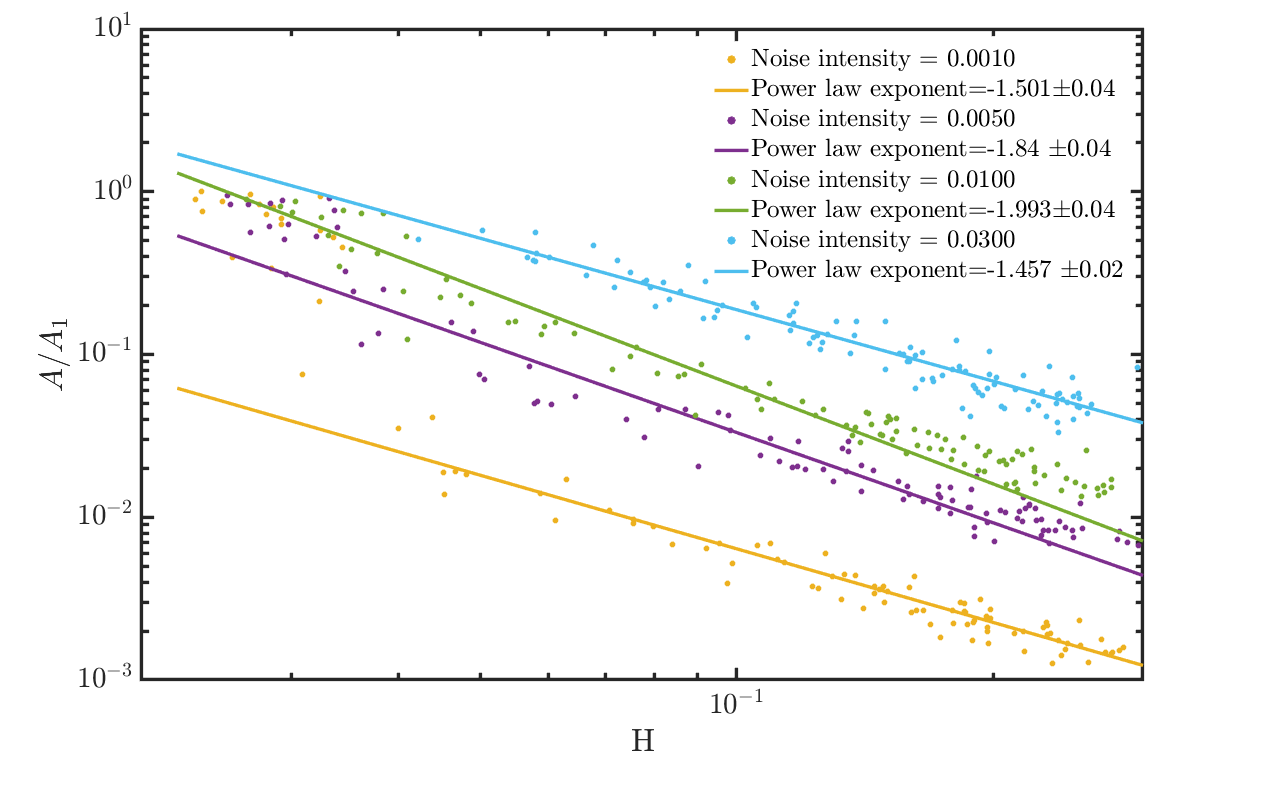}
    \caption{The inverse power law relation of the normalized spectral amplitude ($A/A_1$) of the dominant frequency with the Hurst exponent $(H)$ as the control parameter $\mu$ is increased from -0.25 to 0.05. For the noise intensity values of $\eta=0.001$, $0.005$, $0.01$, and $0.03$  the power law exponent with the 95\% confidence interval is obtained as $-1.501 \pm 0.04$, $-1.84 \pm 0.04$, $-1.993 \pm 0.04$, and $-1.457 \pm 0.02$, respectively.}
    \label{fig: differentq scaling}
\end{figure}

The degree of self similarity for a multifractal system exhibits spatiotemporal variation. As a result, several fractal dimensions are needed to accurately characterize a multifractal system. The Hölder exponent $\alpha$ charaterizes the local singularity strength, which quantifies the degree of regularity and irregularity in the system's behavior at a point. The multifractal spectrum ($f(\alpha)$) describes the fractal dimension of the set of points in the system that share the same singularity strength $\alpha$. 

The estimation of the multifractal spectrum involves first computing the q-order mass exponent, $\tau(q)$, defined as,
\begin{equation}\label{eqn: tau definition}
    \tau(q) = qH(q)-1
\end{equation}
where $H(q)$ is the generalized Hurst exponent. The Hölder exponent ($\alpha$) is then obtained as the derivative of $\tau(q)$ with respect to $q$,
\begin{equation}\label{eqn: singularity exponent definition}
    \alpha=\tau'(q)
\end{equation}
The corresponding singularity dimension, $f(\alpha)$, is computed via a legendre transform as,
\begin{equation} \label{eqn: singularity dimension definition}
    f(\alpha)=q\alpha -\tau(q)
\end{equation}
The plot of $f(\alpha)$ vs $\alpha$ refers to the multifractal spectrum.

\section{Hurst exponent scaling for $p'$ model for different noise intensities}\label{App_Hurst_diffnoise}

We analyze the effect of different noise intensities on determining the Hurst exponent for the $p'$ signals obtained from our model. Figure \ref{fig: differentq scaling} shows the different scaling relations obtained between the normalized spectral amplitude and the Hurst exponent. For a noise intensity of $\eta=0.001$, the power law exponent obtained is $-1.501 \pm 0.04$. On increasing the noise intensity, the trends suggest that the power law exponent decreases to $-1.993 \pm 0.04$ for $\eta=0.01$. Upon further increasing the noise intensity the power law exponent starts increasing. The power law exponent attains a value of $-1.457 \pm 0.02$ for the noise intensity of $\eta=0.03$. Therefore, we choose $\eta=0.01$ as the noise intensity for the model to mimic the universal scaling relations observed in experiments. We infer that the signal-noise evolution in experiments is related such that the amplitude of noise is approximately $1\%$ of the amplitude of the nonlinear oscillations in the acoustic pressure.

\newpage
 \bibliography{ref.bib}

\end{document}